\documentclass[amssymb,showpacs, amsmath,aps, prl %, superscriptaddress
]{revtex4}
\usepackage{graphicx}
\usepackage{colordvi}
\newcommand{\PI}{\mathbf P}
\newcommand{\Pf}{\mathbf P'}

\begin{document}

\title{XYZ-polarisation analysis of diffuse magnetic neutron scattering from single crystals}
\author{Werner Schweika}
\affiliation{Institut f\"ur Festk\"orperforschung, Forschungszentrum J\"ulich, D-52425 J\"ulich, Germany}
%\date{25$^{th} Feb. 2010 $}

\begin{abstract}
Studies of diffuse magnetic scattering largely benefit from the use of a multi-detector covering wide scattering angles. Therefore, the different contributions to the diffuse scattering that originate from magnetic, nuclear coherent, and nuclear spin-incoherent scattering can be separated by the so-called XYZ-polarization analysis. 
In the past this method has been successfully applied to the analysis of diffuse scattering by polycrystalline samples of magnetic disordered materials. 
Single crystal studies that exploit the vector properties of spin correlations are of particular interest for furthering our understanding of frustration effects in magnetism.
Based on the symmetry properties of polarised scattering a suitable extension of the conventional XYZ method has been derived, which allows for 
 the complete separation and the  analysis of features of diffuse magnetic scattering from single crystals. 
\pacs{61.12.Bt, 61.12.Ex,75.25.-j,75.25.Dk}
 \end{abstract}

\maketitle

\section{Introduction}
The scattering theory of polarised neutrons  including magnetic interactions 
started  with the early  work of
 by  Halpern and Johnson \cite{halpern}  and  has been essentially completed by   Blume\cite{blume} and Maleyev\cite{maleyev} more than fourty years ago.
The major milestones in development of experimental means  and applications have been set by technique of 
longitudinal polarisation analysis (Moon {\em et al.} \cite{Moon})
and by  spherical neutron polarimetry (Tasset and Brown)  % [5-8]%(\cite{rekveldt,okorokov,tasset,Brown2001}), 
\cite{tasset,Brown2001},
which nowadays provides a high precision tool to analyse the full polarisation tensor for single crystal magnetic scattering.
 The development for diffuse scattering utilizing a wide angle detector has been essentially driven by Sch\"arpf, who established with {\em his} D7 instrument  at the ILL and the so-called XYZ method\cite{xyzTB,xyz}  efficient means  for measuring and analysing spin-correlations
 in powder samples.  With recent instrumental progress, the   D7 at ILL  \cite{Stewart} and the  DNS at JCNS  \cite{ws_dns2000} 
have gained the required efficiency to explore more routinely the diffuse magnetic scattering from single crystals. 
To overcome the pending problem that the XYZ-method is invalid for
the separation of  the more complex scattering from single crystals,
here an appropriate extension  of the XYZ-method 
 is derived using  the symmetry properties of the polarised scattering.
The present study revisits the polarised scattering from single crystals with the useful achievement of  a  proper separation based on the XYZ-method.
that will certainly be useful for exploring the complexity of disorder involving magnetic properties.

\section{Polarised neutron scattering and symmetry}
%\subsection{The Blume-Maleyev equations}
According to Blume \cite{blume} and Maleyev \cite{maleyev}
the neutron scattering process including magnetic interactions can be completely described by two master equations, Eq.(1) for the
scattering cross-section,  here for brevity denoted by the intensity $I$, and Eq.(2) for $\mathbf P' I$, 
where $\mathbf P'$ denotes the final polarization:
\begin{eqnarray} 
 I &=&   N^{\dag} N  +I_{si}  +  \mathbf  M_{\perp} ^{\dag} \mathbf M _{\perp} 
 +  \PI \cdot   \mathbf M_{\perp}  ^{\dag} N
 +  \PI \cdot   \mathbf M_{\perp}   N  ^{\dag}
+ \mathrm i \PI (\mathbf M_{\perp}  ^{\dag} \times  \mathbf M _{\perp} )
\\
 \Pf I &=&  \PI  ( N^{\dag} N   -\frac{1}{3} I_{si}) 
+(\PI \cdot \mathbf M_{\perp}  ^{\dag})\mathbf M_{\perp}  
+(\PI \cdot \mathbf M_{\perp} )\mathbf M_{\perp}   ^{\dag}
- \PI  (\mathbf M_{\perp}  ^{\dag} \mathbf M_{\perp} ) \\ 
&&+\mathrm i N(\PI \times \mathbf M_{\perp}  ^{\dag} )
-\mathrm i N ^{\dag} (\PI \times \mathbf M_{\perp} )
+N \mathbf M_{\perp}  ^{\dag} +N ^{\dag} \mathbf M_{\perp}  
- \mathrm i (\mathbf M_{\perp}  ^{\dag}\times \mathbf M_{\perp} ) \;,
\nonumber
\end{eqnarray}
where
$ %\begin{equation}
  N(\mathbf Q) = \sum_n b_n {\exp}(\mathrm{i} {\bf Q}\cdot {\bf R_n})
$  %\end{equation}
 is the nuclear structure factor and 
$ 
\mathbf M_\perp = \mathbf {e_Q} \times \mathbf M (\mathbf Q) \times \mathbf {e_Q}
$,
is the vector of dipolar magnetic interaction of magnetic moments with the neutron spin, and $ \mathbf {e_Q}$ is
the unit vector along the direction of the scattering vector $\mathbf Q$.
Accordingly, the  parallel components to the scattering vector $\mathbf Q$ do not contribute to magnetic scattering.
Here, 
$ 
\mathbf M (\mathbf Q) =
\sum_n \mathbf M_n {\exp}(\mathrm{i} {\bf Q}\cdot {\bf R_n})
% \int d \mathbf R \, \,  \mathbf M (\mathbf R) \exp (\mathrm{i} \mathbf Q \cdot \mathbf R),
$ 
denotes the Fourier transform of the magnetic moments. 
$I_{si}$ denotes the nuclear spin-incoherent scattering, assuming that the nuclear spins are randomly oriented.
This diffuse background, which is usually small compared to magnetic and nuclear  Bragg peaks,
can be relatively large when considering diffuse magnetic and diffuse nuclear scattering.

The full information about the scattering terms in Eqs.~(1,2) can be retrieved by spherical neutron polarimetry  \cite{Brown2001}. 
Therefore, the most convenient and standard choice of  a $x$,$y$,$z$-coordinate system  for the setting of the  polarisation $\PI$
 is to have one axis $x$ parallel to $\mathbf Q$, with the axes $y$ and $z$ perpendicular to $\mathbf Q$ pointing in- and out-of the scattering plane
 respectively. 
 
 However, when using multi-detectors  $\PI$ can be set ideally parallel to $\mathbf Q$ only for 
a single detector.  Therefore,  the XYZ-polarisation analysis for multi-detectors 
 the axes  $x$ and $y$ are chosen to be in the (horizontal) scattering plane and $z$ to point (vertically) out of the scattering plane.
In an excellent contribution to the present subject, the polarised scattering has been revisited and treated 
within the density matrix formalism by Sch\"arpf \cite{Schaerpf_school}. 
Following Ref.\cite{Schaerpf_school} we consider first the general case  of an arbitrarily rotated  $x$,$y$,$z$-system
and the polarised intensities that can be measured in spin-flip and non-spin flip modes and their relation to the above equations. 
With the definition of polarised intensities and polarisation
\begin{eqnarray}
I=I_{\nu \nu} +I_{ \nu \bar \nu}    \;\;\;\; ,\;\;\;\;  P'_{\nu} I = I_{\nu \nu} - I_{\nu \bar \nu}
\end{eqnarray} 
for any cartesian coordinate $\nu=x,y,z$,  the non-spin flip  (\textit{nsf}) and spin-flip intensities (\textit{sf}),  $I _{\nu \nu }$ and $I _{\nu \bar \nu }$ respectively,
have been derived  \cite{Schaerpf_school} from Eqs.~(1) to (3):
\begin{eqnarray}
 I _{\nu \nu }&=& N^{\dag} N + N M_{\perp \nu }^{\dag} +N^{\dag}  M_{\perp \nu }   + M_{\perp \nu } ^{\dag} M_{\perp \nu } +\frac{1}{3}I_{si}   \\ 
  I _{\nu  \bar \nu}&=&  \mathbf M_{\perp } ^{\dag} \mathbf M_{\perp } - M_{\perp \nu} ^{\dag} M_{\perp \nu} +\mathrm  i  (\mathbf M_{\perp } ^{\dag} \times  \mathbf M_{\perp })_{\nu}  +\frac{2}{3}I_{si}  \;.
\nonumber
\end{eqnarray}
Eq.~4 shows a well known result namely that only components of $\mathbf M_{\perp}$ parallel to the neutron spin appear in the non-spin flip scattering and only  components of $\mathbf M_{\perp}$  perpendicular 
to the neutron spin can contribute to the spin-flip scattering 
and is exemplified for $\nu=x$
  \begin{eqnarray*}
I _{x  \bar x}&=&  M_{\perp y} ^{\dag} M_{\perp y} + M_{\perp z} ^{\dag} M_{\perp z} +\mathrm  i  ( M_{\perp y } ^{\dag}   M_{\perp z} - M_{\perp z } ^{\dag}   M_{\perp y})  +\frac{2}{3}I_{si} \;.
\end{eqnarray*}
It is further worthwhile to  note that with any specific choice of the coordinate system the average of non-spin flip and spin-flip intensities
differ for the different directions
$$I_{x} = \frac{1}{2} (   I_{xx} +I_{x \bar x})       \neq  I_y  \neq  I_z .$$

So far we followed the analysis of Sch\"arpf. For a further analysis and  separation of terms in the polarised neutron scattering, here, 
we continue with separating intensities in symmetric and antisymmetric parts.
Therefore, we consider scattering for reversed polarisation by the intensities  $ I _{\bar \nu  \bar \nu } $  and $  I _{\bar \nu   \nu } $
 which are experimentally likewise accessible, and we obtain the analogue
to Eq.~(4) from  Eqs.~(1-3):
 \begin{eqnarray}
 I _{\bar \nu  \bar \nu }&=& N^{\dag} N - N M_{\perp \nu }^{\dag} -N^{\dag}  M_{\perp \nu }   + M_{\perp \nu } ^{\dag} M_{\perp \nu } +\frac{1}{3}I_{si}  \\ 
 I _{\bar \nu \nu}&=&  \mathbf M_{\perp } ^{\dag} \mathbf M_{\perp } - M_{\perp \nu} ^{\dag} M_{\perp \nu} -\mathrm  i  (\mathbf M_{\perp } ^{\dag} \times  \mathbf M_{\perp })_{\nu}  +\frac{2}{3}I_{si}   \;.
  \nonumber  
 \end{eqnarray}
 In comparison to Eq.~4 and 5, a further simplification is achieved by considering the average non-spin-flip and spin flip intensities, $\Sigma^{nsf}_{\nu}$ and $\Sigma^{sf}_{\nu}$ respectively: 
 \begin{eqnarray}
\Sigma^{nsf}_{\nu} = \frac{1}{2} ( I _{\nu \nu} +  I _{\bar \nu \bar \nu})  &=&  N^{\dag} N  + M_{\perp \nu} ^{\dag} M_{\perp \nu} +\frac{1}{3}I_{si}   \\
\Sigma^{sf}_{\nu} = \frac{1}{2} (  I _{\nu  \bar\nu } +I _{\bar\nu \nu })&=& \mathbf M_{\perp } ^{\dag} \mathbf M_{\perp } - M_{\perp \nu} ^{\dag} M_{\perp \nu} + \frac{2}{3}I_{si}  
\nonumber
 \end{eqnarray}
 showing  the expected property of the unpolarised intensity $I$
\begin{eqnarray}
I=\Sigma_{x} =\Sigma_{y} =\Sigma_{z} = \Sigma^{nsf}_{\nu} +\Sigma^{sf}_{\nu}  
 &=&  N^{\dag} N  + \mathbf M_{\perp } ^{\dag} \mathbf M_{\perp } +I_{si}  \;.  \nonumber
\end{eqnarray}

By this we have eliminated the interference terms related to nuclear-magnetic 
 correlations (spin-orbit coupling) and vector chirality terms (cross products) that now appear in 
 the corresponding averaged deviations $\Delta^{nsf}_{\nu}$ and $\Delta^{sf}_{\nu}$ respectively, :
  \begin{eqnarray}
\Delta^{nsf}_{\nu}=\frac{1}{2} ( I _{\nu \nu} -  I _{\bar \nu  \bar\nu }) 
&=&N M_{\perp \nu }^{\dag} +N^{\dag}  M_{\perp \nu }   =2 \Re( N M_{\perp \nu }^{\dag})\\
\Delta^{sf}_{\nu } =
 \frac{1}{2} (  I _{\nu   \bar \nu } - I _{\bar \nu  \nu })
 &=& 2 \mathrm i (\mathbf M_{\perp } ^{\dag} \times  \mathbf M_{\perp })_\nu  \;.   \nonumber 
\end{eqnarray}
Eq.~(7) reveals a well known result that is the basis of  experiments with polarized neutrons using polarisation reversal
without polarisation analysis.
 For the purpose of separating the interference terms due to chirality or due to nuclear-magnetic 
 correlations, polarisation analysis is actually
not required and can be achieved by setting $\nu$, with polarisation reversal, either parallel (for chirality) or perpendicular (for nuclear-magnetic interference)
to the scattering vector $\mathbf Q$.
However, as can be seen, Eq.~(7) also provides a complete separation of these terms for any choice of the cartesian coordinate system
if  polarisation analysis is applied.

Finally, one may note that it is also straightforward 
to derive this symmetry decomposition from Brown's tensor representation of polarised scattering \cite{Brown2001}
for the case of the standard coordinate system ($x || \mathbf Q$) and also to include spin-incoherent scattering.

\section{XYZ-polarisation analysis, powders and single crystals}

The standard XYZ-method has been successfully applied in many studies of magnetic correlations using powder samples.
From the directional dependence of the polarised scattering, the purely magnetic scattering contribution can be separated 
by experimental means of polarisation analysis. For further interest the reader is referred to a recent and detailed review \cite{Stewart}
and a recent own study \cite{kagome}. 
 Here we shall briefly recall the fundamental equations (Eq.~8), and discuss the additional potential of instruments like DNS and D7 for 
studies of powder samples with polarisation analysis.  
An interesting consequence is that the interference of nuclear and 
induced magnetic scattering could be observed even for powders if chiral domains relate to satellites with different $Q$ moduli or in presence of  a symmetry breaking external field, whereas
usually,  chiral scattering is found in  single crystals with a preferred domain orientation.
In the following (3.2) 
the central issue will be  to separate and analyse the different contributions  to the polarised  scattering from single crystals
based on the conventional tools of the standard XYZ method. 
Hence, the separation will be achieved  without approximations only from diagonal intensity elements.

 \subsection{ The application to powders}
In case that orientational averaging applies, e. g. for powder samples,  
we do not distinguish the components $M_{\perp \nu}$, 
and the magnetic scattering can be separated 
via the XYZ polarisation analysis \cite{xyzTB,xyz} % (Sch\"arpf and Capellmann, 1993)
 from  the diagonal intensities.
In case of only weak guide fields 
one does not need to distinguish the intensities with respect  
to the sign of polarisation

\noindent
\raisebox{0mm}{\hspace{0mm}\begin{minipage}[b]{33pc}
as proposed in the previous section. It is a particular virtue of the XYZ-method  that this 
method applies  even if the Cartesian coordination system 
is arbitrarily rotated in the scattering plane, so that the new coordinates are $x',y',z$
(see figure). 
In case of multi-detectors covering a larger Q-range it is trivially impossible to have the polarisation simultaneously  parallel to all different Q vectors; 
the difference between the polarisation axis $x'$ and the actual $\mathbf Q$ vector  
is the \textit{Sch\"arpf angle}  $\alpha$. 
\end{minipage}}\hspace{1pc}\includegraphics[width=5.8pc,angle=90]{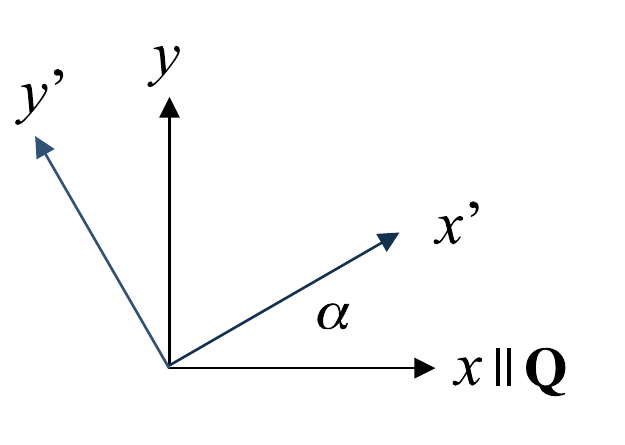} 

The information about the in-plane magnetic intensities, the square of magnetic amplitudes 
  $\mathbf M _{\perp y}$,
is recovered from 
$   \mathbf M _{\perp y}^2= \mathbf M _{\perp x'}^2+ \mathbf M _{\perp y'}^2  =  \mathbf M _{\perp y}^2 (\sin^2 \alpha  +   \cos^2 \alpha)  $. 
Hence, the magnetic, nuclear coherent and spin-incoherent scattering can be separated by combinations of polarised intensities \cite{xyzTB,xyz,Stewart}
%\begin{figure}[h]
%\includegraphics[width=5pc,angle=90]{alpha.eps} %\hspace{1pc}
%\begin{minipage}[b]{30pc}
\begin{eqnarray}
|\mathbf M_\perp|^2   %\MM
&=& 2(I_{x'\bar x'}+I_{y'\bar y'}-2I_{z\bar z})^{sf}
 = - 2(I_{x' x'}+I_{y' y'}-2I_{z z})^{nsf}
\\
 I_{si} & =& {3\over 2} \left(
-I_{x'\bar x'} -I_{y'\bar y'}+   3 I_{z\bar z}
 \right)   \nonumber
\\ N^2 &=& I_{z z} -\frac{1}{2}  |\mathbf M_\perp|^2 %\MM 
- \frac{1}{3} I_{si}  \nonumber 
\end{eqnarray}
%\end{minipage}
%\end{figure}
Here, we have assumed that the intensities do not change for a common polarisation reversal of initial and final polarisation.
More carefully, for the application of this analysis we have to consider whether the antisymmetric part actually vanishes. 
This is fulfilled only if there is no distinguished direction by any external field. Guide fields applied in the XZY method are
typically in the order of magnitude of ten Gauss and they are too weak to induce any significant  sample magnetization in paramagnetic
or antiferromagnetic samples.  

Strong fields invalidate the standard XYZ separation method, however, this allows for the study of
the correlations and interference terms between nuclear and magnetic scattering amplitudes by determining the antisymmetric 
part of the polarised intensity, \textit{id est} measuring intensity differences for polarisation reversal. A unique response of chiral terms to
an applied horizontal external field is less likely, although  not impossible.
A well-known strategy  for single crystals used in a ``half-polarised"  setup
 \cite{schweitzer}  
can also be applied to powder diffraction. 
Such a determination of correlations between nuclear and magnetic scattering amplitudes  from powders looks promising 
particularly for detectors covering a large solid angle without polarisation analysis.
With respect to polarisation analysis of scattering it 
 may be advantageous that only the difference of the \textit{nsf}-intensities contributes to the signal, 
 and the  background from \textit{sf}-scattering could be eliminated. % using fully polarised instruments. % like DNS or D7 .  

\subsection{The application to single crystals}
For a single crystal, even without applied external field, its intrinsic anisotropy and possible polarity 
may give rise to antisymmetric scattering contributions. 
Therefore, the symmetry decomposition resulting in Eqs.~(6,7)
is important and helpful to analyse the polarised scattering and its magnetic contributions. 
For a Cartesian coordinate system ($x',y',z$) rotated
by the angle $\alpha$ around the vertical $z$-axis with respect to 
 $x$ parallel to $\mathbf Q$,
  the following relations can be derived  from Eqs.~(6) and (7) providing a complete separation 
  \begin{eqnarray}
%N^2 
N^{\dag} N&=& \frac{1}{2} \left( \Sigma^{nsf}_{x}+\Sigma^{nsf}_{y} -\Sigma^{sf}_{z}  \right) 
= \frac{1}{2} ( \Sigma^{nsf}_{x'}+\Sigma^{nsf}_{y'} -\Sigma^{sf}_{z}  )   %\nonumber 
\\
I_{si} &= &\frac{3}{2} 
( \Sigma^{nsf}_{x}-\Sigma^{nsf}_{y} +\Sigma^{sf}_{z}  ) = 
\frac{3}{2}  
%\left( 
\frac{ \Sigma^{nsf}_{x'}-\Sigma^{nsf}_{y'}} 
{\cos ^2 \alpha - \sin^2 \alpha }+
% {1-2 \sin ^2 \alpha}+
%
\frac{3}{2} 
 \Sigma^{sf}_{z}  
 \\
M_{\perp y} ^{\dag} M_{\perp y}  
&=&\Sigma^{sf}_{z} -  \frac{2}{3}I_{si}   %\nonumber 
\\
M_{\perp z} ^{\dag} M_{\perp z} 
 &=&\Sigma^{nsf}_{z} -  \frac{1}{3}I_{si} -  %N^2 
 N^{\dag} N %\nonumber
 \\
  I_{chiral,x'}&=& % \frac{1}{2} (  I _{x   \bar x } - I _{\bar x  x })
 2 \mathrm i (\mathbf M_{\perp } ^{\dag} \times  \mathbf M_{\perp })_{x'}  % \nonumber %
 =\Delta^{sf}_{x'}        \;\;=\cos \alpha \, \Delta^{sf}_{x} 	%\nonumber 
 \\
I_{NM,y'}&=&2 \Re( N Q_{y'}^{\dag})   \, \; \;\;\;\;\;\;\; \;   = \Delta^{nsf}_{y'}  =\cos \alpha  \, \Delta^{nsf}_{y}  %\nonumber
 \\
I_{NM,z'}&=&2 \Re( N Q_{z'}^{\dag})   \,  \;\;\;\;\;\;\;\; \;= \Delta^{nsf}_{z'} =   \;\; \;\;\;\;\;\;\,  \Delta^{nsf}_{z}   \;, % \nonumber 
\end{eqnarray}
where the geometrical correction factors in Eqs.~(10,13,14) stem from the projections $M_{\perp x'}=\sin \alpha  \,M_{\perp y}$  and $M_{\perp y'}= \cos \alpha  \,M_{\perp y} $  
to the scattering vector, see comment \cite{footnote}.

Compared to the information content of powder experiments, the study of diffuse polarised scattering from single crystals 
yields additionally the anisotropy of magnetic correlation functions, possible vector chirality  and all nuclear- magnetic correlations,
essential information about  complex magnetic materials of high current interest. 
It is worthwhile to  note that a separation of the chiral term has already been given earlier by Eq.~50 in Ref.\cite{xyz}.

For diffuse scattering 
 the typical experimental strategy is to map out scattering planes by  stepwise rotations of the crystal around the vertical axis $z$.
 Note, the separation by Eqs.~(9) to (15) is independent of the sample rotation $\omega$ and 
  the convenient choice is a fixed coordinate system  $x',y',z$. 
  
  In order to interpret the magnetic scattering amplitudes in terms of the crystal lattice, 
 the in-plane component $M_{\perp  y} $ has to be decomposed further into independent lattice 
 components in the plane (ab) and is given by the projections of  $M_{a} $ and $M_{b} $, while the 
 vertical component is simply identified   by $M_{c} $.  For orthogonal components, only the squares of $M_{a} $ and $M_{b} $ 
 determine  $M_{\perp y} $, see Fig.~1. Note this example demonstrates that the considerations here also apply to inelastic scattering.
 \begin{figure}[h]
\includegraphics[width=12.9pc,
angle=90]{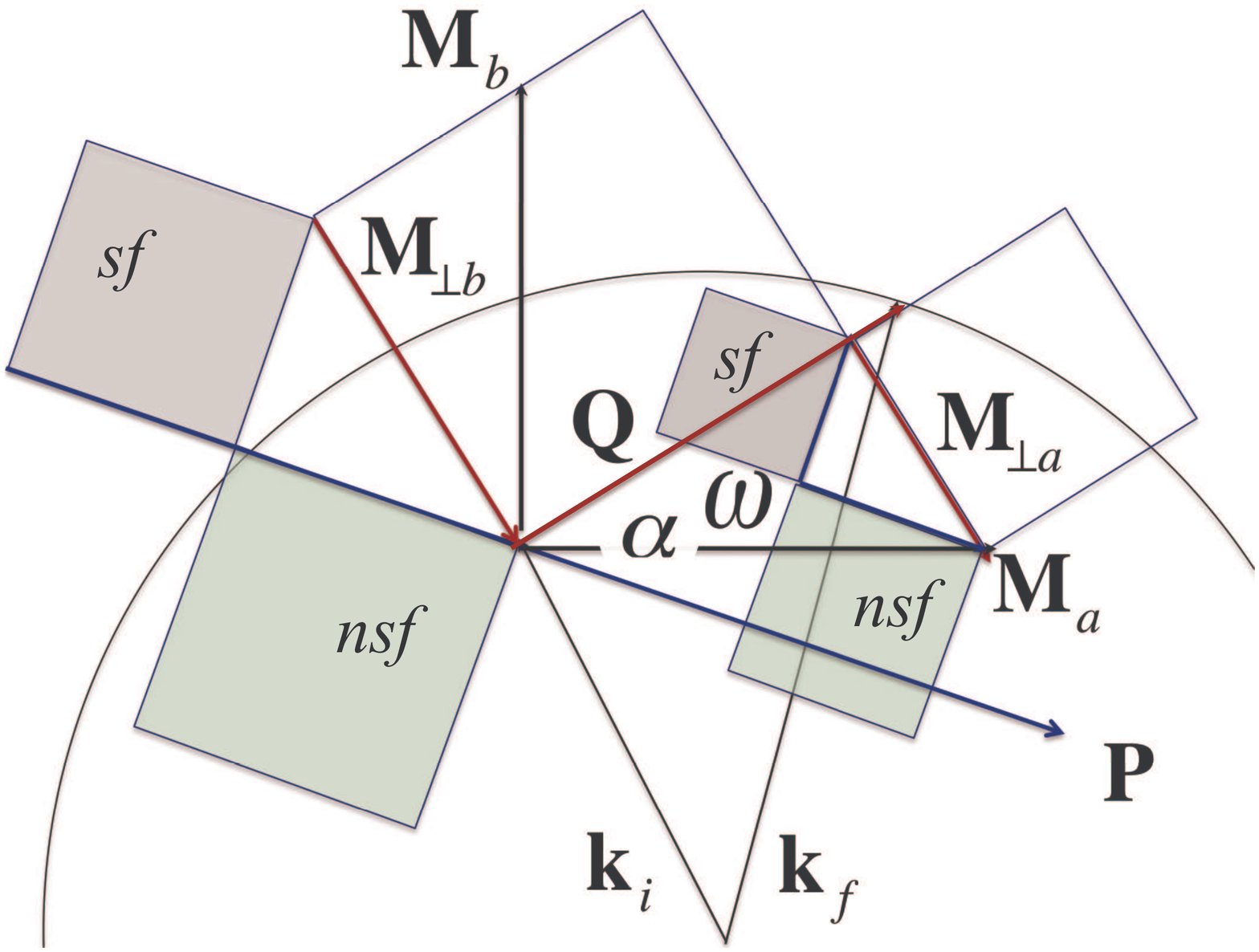}\hspace{2pc}%
%\begin{minipage}[b]{23pc}
\caption{\label{label} (color online) Polarisation $\mathbf P$ and magnetic (inelastic) scattering $M_{\perp \nu}^2$ (white squares) for the case of a 
\textit{Sch\"arpf angle} $\alpha$ between scattering vector $\mathbf Q$ and polarisation $\mathbf P$; 
here $\omega$ denotes the angle between  $\mathbf Q$ and reciprocal lattice vector $\mathbf a^*$ with $\mathbf M_a || \mathbf a^*$. Shaded squares represent \textit{sf}- and \textit{nsf}-intensities related to the projections of $\mathbf M_{\perp a}$ and $\mathbf M_{\perp b}$ $\perp$ and $|| \mathbf P$ respectively. The vertical component ($||z$ and $\perp$ to $\mathbf P$, \textit{sf}-intensity)  is not shown.}
%\end{minipage}
\end{figure}
\vspace*{-0.3mm}
 The disentanglement of $M_{a} $ and $M_{b} $ is, analog to structure determination or to disorder in displacements variables, 
 relying on sufficient information in extended Brillouin zones of reciprocal space.  
Diffuse scattering related to disorder phenomena typically implies an exponential decay of correlations in real space.
Hence a Fourier analysis will efficiently reveal finite real space pair correlation functions,  an approach that is 
a linear least squares problem having a unique solution.  Therefore, the determination of  three dimensional pair-correlations functions
requires accordingly measurements of more than a single scattering plane. 

\section{Final remarks}
With recent instrumental developments,  particularly on the DNS and D7, diffuse magnetic scattering of single crystals can be efficiently measured  
with polarisation analysis, 
which has already provided exiting results on frustrated magnetism in spin-ice compounds \cite{Fennell}; a further example can be found in the proceedings of the PNSXM 2009 \cite{Chang}.
 For a systematic and thorough understanding the separation of the different components is essential, a task that  to date 
 has been believed to be incapable by the conventional means used in XYZ polarisation analysis. 
 The solution here, utilizing the symmetry properties,  also places the emphasis on additional measurements with polarisation reversal. 
 It is worthwhile to note that this request does not demand for more beamtime to achieve equal statistical weights. 
 
 So far we have not considered the off-diagonal terms. 
 These terms also represent the interference terms, Eqs.~(13-15), due to chiral correlations and due to coupling of spin-space 
 and real-space variables.
 Spherical neutron polarimetry of off-diagonal terms distinguishes 
 these effects irrespectively of  a possible depolarisation from different magnetic domains in the crystal. 
 Obviously, the diagonal antisymmetric polarised intensities Eq.~(7) also reveal the mentioned interference terms.
It should be noted that  the same XYZ-setup for multi-detectors can be used to access the complete polarisation tensor 
and off-diagonal terms can be determined with the same accuracy by  spherical neutron polarimetry
 with precessing incident polarization\cite{ws_vector,ws_nn}. 
 However,  the here proposed XYZ-polarisation analysis, a method based on only diagonal terms and  applicable for multi-detectors, 
will certainly provide a much more convenient  and efficient separation for the diffuse magnetic neutron scattering from single crystals.

\section*{Acknowledgments}
Otto Sch\"arpf is gratefully acknowlegded for discussions and valuable comments.

\end{document}